\documentclass[aps,prl,twocolumn,groupedaddress,showpacs,nofootinbib]{revtex4}

\usepackage{bm}
\usepackage{graphicx}
\usepackage{times}
\usepackage{amssymb,amsmath}

\def\jcap{JCAP}
\def\aj{AJ}
\def\apj{ApJ}
\def\apjl{ApJL}
\def\prl{PRL}
\def\prd{PRD}
\def\jhep{JHEP}
\def\pl{Phys.Lett.}
\def\beq{\begin{equation}}
\def\eeq{\end{equation}}
\def\bey{\begin{eqnarray}}
\def\eey{\end{eqnarray}}
\def\bfig{\begin{figure}}
\def\efig{\end{figure}}

\def\lsim{\mathrel{\raise.3ex\hbox{$<$\kern-.75em\lower1ex\hbox{$\sim$}}}}
\def\gsim{\mathrel{\raise.3ex\hbox{$>$\kern-.75em\lower1ex\hbox{$\sim$}}}}

\def\txt{\text}

\def\mpl{\txt{m}_{\txt{pl}}}

\def\order{\mathcal{O}}

\def\H{\text{H}}
\def\pdir{\partial}
\begin{document}
\title{Topology and Dark Energy: Testing Gravity in Voids}

\author{Douglas Spolyar$^1$, Martin Sahl\'en$^{2}$, and Joe Silk$^{1-3}$}
\affiliation{Institut d'Astrophysique de Paris - 98 bis boulevard Arago - 75014 Paris, France$^{1}$
\\BIPAC, Department of Physics, University of Oxford, Denys Wilkinson Building, 1 Keble Road, Oxford OX1 3RH, UK$^2$\\
The Johns Hopkins University, Department of Physics \& Astronomy, 3400 N. Charles St., Baltimore, MD 21218, USA$^3$
}

\begin{abstract}
Modified gravity has garnered interest as a backstop against dark matter
and dark energy (DE). As one possible modification, the graviton can become massive, which introduces a new scalar field - here with a galileon-type symmetry. The field can lead to a nontrivial equation of state (EOS) of DE which is density-and-scale-dependent. Tension between Type Ia supernovae and {\it Planck} could be reduced. In voids the scalar field dramatically alters the EOS of DE, induces a soon-observable gravitational slip between the two metric potentials, and develops a topological defect (domain wall) due to a nontrivial vacuum structure for the field. 
\end{abstract}


\maketitle

{\bf  INTRODUCTION}
{
The non-detection of dark matter and the incomprehensibility of dark energy are challenging puzzles for theorists.}
Alternatives to General Relativity (GR) is one strategy that is useful to have should the current dark sector crisis persist.
In modifying gravity, the one lesson that has been learnt is the difficulty 
of finding an alternative theory that satisfies experimental constraints on all scales where gravity can be probed. 
High precision measurements on laboratory and solar system scales must accommodate  
astrophysical constraints on the two metric potentials in modified gravity theory
 from galactic to cosmological scales.  Growth of density fluctuations  and galaxy dynamics constrain the Newtonian potential.  
Light propagation constrains the sum of the two potentials. See~\cite{2010AnPhy.325.1479J,Clifton:2011jh,Hinterbichler:2011tt} for review and citations.	

Modifying gravity  generally introduces a scalar that complements and couples to  the tensor field of general relativity. 
The scalar must however be screened on small scales to be
 reconciled with the local precision tests of gravity. Three main mechanisms have been discussed for screening the new field. 
 Chameleon~\cite{Khoury:2003aq} and f(R) theories~\cite{1970MNRAS.150....1B}  screen via density and decouple the scalar field at high 
 density. Symmetron fields~\cite{Hinterbichler:2010es} weaken the scalar coupling at high density. 
 Galileon/Braneworld fields~\cite{2009PhRvD..79f4036N,Dvali:2000hr} and massive gravity (focus of this paper) operate via the Vainshtein effect~\cite{Vainshtein:1972sx},
  which decouples the scalar field in high-curvature regions. 
 
Massive gravity (by assigning the graviton a mass of order the inverse Hubble scale) naturally and transparently restores coupling on the horizon scale, where one may then hope to account for the 
accelerated expansion of the universe~\cite{2011PhRvD..83j3516D}. The transition scale (i.e.  the Vainshtein radius where decoupling occurs) is of order tens of Mpc today. This scale corresponds to the largest self-gravitating scales in the universe, and potentially the simplest to understand. {
 While massive gravity theories offer a rich phenomenology and resolve some theoretical issues, there are also remaining theoretical difficulties, e.g. that quantum corrections may render theories unreliable on small scales \cite{2013PhRvL.111b1802B}, or possible issues with acausality \cite{2013PhRvL.110k1101D}.}

In galaxy clusters, the metric potentials  are almost the same as with GR, which is not 
necessarily true in voids. The change in the potential is shown here to only be a few percent for over-densities (which has been found previously by \cite{Wyman:2011mp,2013JHEP...02..080S,2012PhLB..713...99C}) but of order unity for voids.  The apparent change in dynamics and lensing can be parameterized as an inhomogeneous (scale-dependent) EOS of DE, which could be measured with future galaxy surveys. Even more surprising, the scalar field forms a topological defect (domain wall) at the edge of voids.

\begin{figure}[t]
\includegraphics[width=.5\textwidth]{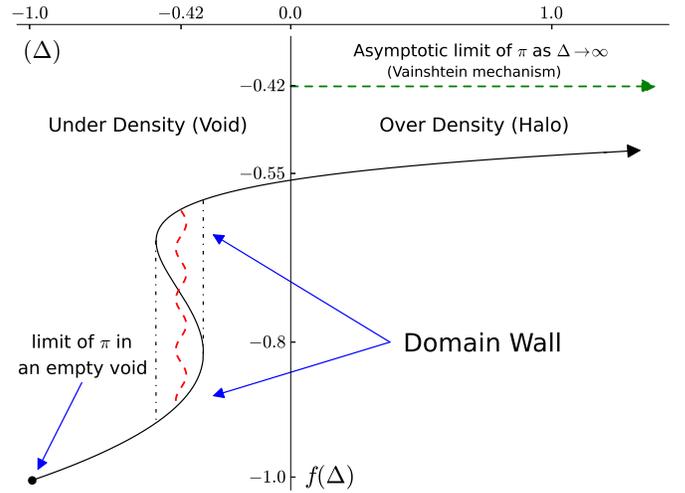}
\caption{ \label{piprime}Plot of $f$ with $\pi^\prime/r= f(\Delta) \Lambda^3$ where $\Delta=\delta\rho_{\rm m}/\rho_{m0}$ is the difference between the mean density inside of a radius $r$ and the mean density today normalized by the mean density today. {
The vertical black dashed lines indicate the domain wall region. The wiggly red dashed line is meant to indicate that a field transition between branches takes place within the region.}}
\vspace{-17pt}
\end{figure}

{\bf MASSIVE GRAVITY}
Massive gravity (linearized) corresponds to a spin-2 field with
 five degrees of freedom, which can be decomposed into tensor, vector, and scalar modes.
The effect of massive gravity in voids is 
well-characterized in the weak-field limit of GR and on sub-horizon scales, which is exactly the decoupling limit (see~\cite{2009PhRvD..79f4036N,2010PhLB..693..334D,2010PhRvD..82d4020D,2013JHEP...02..080S} for a precise definition). 
In this case, an effective theory  characterizes massive gravity, which only depends upon the tensor and scalar degrees of freedom. The vector
degrees of freedom decouple leaving a simple Lagrangian  given by \cite{2011PhRvD..83j3516D},
\beq
\label{lagrangian}
\mathcal{L}=
h^{\mu\nu}\bigg(-\frac{1}{2}\mathcal{E}^{\alpha\beta}_{\mu\nu}h_{\alpha\beta}+\frac{1}{\mpl}\txt{T}_{\mu\nu}+\mathcal{L}(\pi)_{\mu\nu}\bigg).
\eeq
The Einstein tensor $(\mathcal{E}_{\mu\nu}^{\alpha\beta}h_{\alpha\beta})$ comes from the Einstein--Hilbert Lagrangian (m$_{\txt{pl}}^2\sqrt{-\txt{g}}$ R)
 assessed to the second order in metric fluctuations $h^{\mu\nu}$ with $g_{\mu\nu}=\eta_{\mu\nu}+h^{\mu\nu}/\mpl$,  the Newton constant $8\pi$G~$\equiv \mpl^{-2}$,
 $\eta_{\mu\nu}=(-1,1,1,1)$, and $\mpl$ is the Planck mass.
 The second term of Eq.~\ref{lagrangian} is the standard coupling of GR to the stress-energy tensor. 
 The third term is the Lagrangian for the scalar field $\pi$ with $\mathcal{L}(\pi)_{\mu\nu}=\alpha X^{(1)}_{\mu\nu}+(\beta/\Lambda^3) X^{(2)}_{\mu\nu} 
 +(\gamma/\Lambda^6) X^{(3)}_{\mu\nu}$, where $\alpha, \beta,$ and $\gamma$ are arbitrary parameters of the theory but are taken to be $\order(1)$.
The effective field theory for $\pi$ is valid up to the cutoff scale $\Lambda^3=\mpl m^2$ ($m$ is the graviton mass). See \cite{2011PhRvD..83j3516D,2010PhRvD..82d4020D} for  details.

 {\bf Ansatz}
 We look for solutions which can accommodate both under- and over-densities (e.g.~voids, and halos/solar system respectively) and are consistent with cosmology. 
 The  equivalence principle implies that locally we can always define a metric (for all time) which will simply be Minkowski space 
 plus a perturbation.  The perturbation component will account for  the non-trivial geometry of the space-time. Such coordinates are referred to as Fermi coordinates~\cite{2009PhRvD..79f4036N,Manasse:1963zz}.
  In the Newtonian gauge,
\beq
 \label{ansatz}
 {\rm d}s^2=-(1+2\Phi(r)){\rm d}t^2+(1-2\Psi(r)){\rm d}x^2.
\eeq
 which on small scales can be deduced from the general FRW metric by using the transformations $t_{\rm c}=t+\H^2\vec{x}^2/2$
  and $x_{\rm c}={\vec{x}}/{a}(1+1/4 \H^2\vec{x}^2)$
  between co-moving ($t_{\rm c}, x_{\rm c}$) and the local, physical Fermi coordinates $(t,x)$. $\H$ is the Hubble parameter, and $a$ is the scale factor.
  We then add on local perturbations of the metric. $\Phi$ and $\Psi$ encode both the background geometry and local metric perturbations.
The standard stress--energy tensor (in Fermi coordinates)  is simply $T_{\mu\nu}\simeq(\rho,\delta_{ij}p)$ where we 
have neglected off-diagonal terms ($\order(\H^2x^2)\sim v^2\sim\H^2 x^2v^2$ with velocity $v$). 
Similarly, $\rho$ and $p$ include local perturbations of matter and radiation relative to a background density.
Finally, we make the ansatz that the field $\pi$ (as well as the local metric and matter perturbations) are spherically symmetric and time-independent, $\pi=\pi(r)$. 
{
 A subset of solutions with the above ansatz has been explored in Refs. \cite{2012PhLB..713...99C,2013JHEP...02..080S} - we find and explore a new class of solutions.}

By the least-action principle applied to Eq.\ref{lagrangian}, we obtain two
non-trivial  Equations of Motion (EOM) for the metric and one EOM for  $\pi$ with $\prime={\rm d}/{\rm d}r$
\beq
\label{metric1}
2\nabla^2\Psi=8\pi G \langle\rho\rangle+\frac{6\,\alpha}{\mpl}\frac{\pi^\prime}{r}+\frac{6\,\beta}{\mpl \Lambda^3} \frac{(\pi^\prime)^2}{r^2}+\frac{6\,\gamma}{\mpl \Lambda^6} \frac{(\pi^\prime)^3}{r^3}
\eeq
\beq
\label{metric2}
(\nabla^2-\pdir_i^2)(\Phi-\Psi)=8\pi G \langle p\rangle-\frac{4\,\alpha}{\mpl}\frac{\pi^\prime}{r}-\frac{2\,\beta}{\mpl\Lambda^3}\frac{(\pi^\prime)^2}{r^2}
\eeq
\beq
\label{piEOM}
\alpha\nabla^2(2\Psi-\Phi)+\frac{2\,\beta}{\Lambda^3} \frac{\pi^\prime}{r}\nabla^2(\Psi-\Phi)-\frac{3\,\gamma}{\Lambda^6}\frac{(\pi^\prime)^2}{r^2}\nabla^2\Phi=0
\eeq
where in Eq.~\ref{metric2} we are not summing over $i$ but indexing over the spatial dimensions ($x,y,z$). 
  Previous authors~ \cite{2013JHEP...02..080S,2012PhLB..713...99C}
have found the above EOM, but have neglected the effect of pressure in Eq.~\ref{metric2}. 
   The operator $\langle\,\,\rangle$ gives  the average value of a given quantity inside a radius $r$.

In linearized GR, perturbations and the background can be separated by subtracting off the background 
from the perturbed Einstein equations. In massive gravity, this proves impossible.  Eq.~\ref{piEOM} is non-linear in $\pi$ and in the metric perturbations, which introduces cross-terms between the 
perturbed solution and the background solution.
At zeroth order in $\pi$ we look for solutions that accommodate both the background and local perturbations in the metric, density, and pressure.

The above EOM can be made dimensionless by multiplying by  $m^{-2}$ and then setting $\pi^\prime/r=f(\Delta)\,\Lambda^3$, where $\Delta=\delta\rho_{\rm m}/\rho_{m0}$.  Here, $\rho_{m0}$ is the matter density today and $\delta\rho_{\rm m}$ is the under/over-density 
of a void/halo with $\delta\rho_{\rm m}(r)~=~\langle\rho_{\rm m}(r)\rangle-\rho_{m0}(r)$.  Hence, $\delta\rho_{\rm m}/\rho_{m0}=-1$ corresponds to an empty void as in devoid of matter but not DE.
 After making the appropriate substitutions, Eq.~\ref{piEOM} becomes a quintic constraint equation for $f(\Delta)$ which  depends
upon the average pressure $\langle p\rangle$ and density $\langle\rho\rangle$ for a fixed radius $r$. Out of the five roots, three will typically be real. 
 
{\bf Classes of Solutions}
The solutions for $\pi^\prime$ can be categorized  into three separate classes. In a separate publication we will discuss in detail the various solutions.
Two of the classes have been
discussed previously by~ \cite{2013JHEP...02..080S,2012PhLB..713...99C}.  Neither of these cases produce interesting cosmological solutions. One class of solutions degravitates all mass. The second set of solutions
generates an EOS of the universe which is equivalent to radiation at late times and generates negative densities.  

The third class gives new cosmological solutions.  The EOS of DE
  tracks the energy density of matter, radiation, and curvature with several different sub-cases.  In the first self-accelerating case [no Cosmological Constant (CC)],
 we can tune $\alpha,\beta,$ and $\gamma$ to effectively 
generate a CC. With a different ansatz, the authors of \cite{2011PhRvD..83j3516D} found a similar solution which decoupled  the scalar field
from matter, radiation, and the metric. In a second self-accelerating solution (flat background),
the EOS of DE will depend on the local density of matter.  In a third self-accelerating case (with curvature), the scalar field can dynamically 
counteract the effect of curvature upon the expansion rate of the universe, which  may appear as so-called ``phantom" EOS of DE with $w<-1$. 

The more general case with a CC will constitute the focus of the paper.  
In solving for the  $\pi$'s EOM, we search  for  roots of a quintic equation which has no general analytic solution. 
We find solutions numerically, such that densities are positive, the field $\pi$ is ghost-free, 
and constraints on $\H_0$ and $\Omega_{\rm m}$ are satisfied.  To be concise {
we will consider two cases with $\alpha= -0.45\,(-0.50)$, $\beta = -1.0\,(-1.07)$, and $\gamma = -0.86 \,(-0.87)$ in which a topological feature appears (for other choices of parameters no domain wall exists).
We set $8\pi G\rho_0/m^2=7.65\,(18.5)$ in order to have the correct matter density today 
where $\rho_0$ is the critical density.}
In Fig.~1, we have plotted the solution of $\pi^\prime/r\Lambda^3=f(\Delta)$ (for case 1) interpolating between an under-density and an over-density. 
 {
 We note that the qualitative behavior of the two cases (with different $\alpha,\beta,\gamma$) are the same.}

{\bf Over-Density }
First, we  consider a point mass $M>0$ (as our over-density).  In the limit of approaching 
the point mass $r\rightarrow 0$, then $\Delta\rightarrow \infty$ and  $f(\Delta)\rightarrow -0.42$. The Vainshtein mechanism kicks in recovering 
GR  ($\Phi\rightarrow\Psi\rightarrow-G M/r$).  As $r\rightarrow\infty$ then we smoothly go onto the cosmological solution and $f(\Delta)$  takes on a value which 
gives the correct cosmological solution today with $2\Phi\rightarrow-(\dot{\H}+\H^2)\vec{x}^2$ and $2\Psi\rightarrow\H^2\vec{x}^2$ where
H is the Hubble parameter.  From the metric EOM, we then also recover the Friedmann equations.

{\bf Under-Density}
 Far outside the void $\delta\rho_{\rm m}\simeq0$ (the metric potentials match onto the cosmological solution found in the previous section).
The average density $\langle\rho_{\rm m}\rangle$ drops as we move to the center of the void.
Eventually  the derivative
of $\pi$ must jump to a new branch (see Fig.~1)  once $\delta\rho_{\rm m}$ drops to less than half of the present mass density of the universe today.  
The jump forces a discontinuity in the derivative of the field $\pi^\prime$. We still have the freedom
to adjust the integration constant of $\pi^\prime$ such that $\pi$ is continuous. Regardless, to match the derivative across the discontinuity requires the introduction of a wall. 

The nature of the $\pi$ wall is categorically  
different from a wall formed for a typical scalar field. The wall is not formed primordially but forms dynamically at late times. Typically, a domain wall  corresponds to a kink solution interpolating between two different  
vacuum states on different sides of the wall. 
 The wall tension is due to the potential separating the two  vacuum states in field space.  
The $\pi$ field has no potential, $V(\pi)=0$, but 
upon removing the static condition, the EOM for $\pi$ becomes highly non-linear. 
As a parallel in fluid mechanics, non-linearities of the Navier--Stokes 
equation can lead to a discontinuity in density and pressure gradients {
(a shockwave). The fluid mechanic description breaks down  and microphysics kicks in which removes the discontinuity.
 Similarly, we conjecture that the full non-linear EOM for $\pi$  (in the presence of matter~\cite{Endlich:2010zj})
will generate a ``shockwave."  At the discontinuity, the effective field theory for $\pi$ breaks down and new physics kicks in which resolves the discontinuity.}

From Fig.~1, there exists a mapping of $\pi^\prime/r$  and $\pi$ into real space, since there is a one-to-one mapping between $\pi$ and $\pi^\prime$.
 The mapping of $\pi^\prime$ is not one-to-one into real space.
 If we allow a complex extension of $\Delta$ and $\pi^\prime$, 
  the quintic equation corresponds to a hyper-elliptic surfaces (a compact Riemann surfaces with genus greater than one), 
 with multiple branch cuts.
$f$ loops around two branch points producing the multivalued function in real space in Fig.~1 but  $f$ is single valued on the Reimann surface!
The $\pi$ field appears to be a NGB (see \cite{2012JHEP...06..004G}).
Frequently, the vacuum structure of a Nambu--Goldstone Boson (NGB) is mapped into real space. By analogy,
we identify Riemann surface as the vacuum of  $\pi$.

We note that the topologically non-trivial vacuum of a NGB when mapped into real space leads to the appearance of topological structures.
   Similarly, our function $f$ due to its wrapping around branch points is equivalent to  a non-trivial wrapping
around the hyper-elliptic surface. 
Finally, if our solutions for  $\pi$ are `onto', then the non-renormalization theorem for $\alpha,\beta,\gamma$ may be a consequence of the compact 
vacuum of $\pi$, since the terms in $\mathcal{L}(\pi)$ are in fact Wess--Zumino--Witten (WZW) terms~\cite{2012JHEP...06..004G}.
Hence further properties of the field may potentially be inferred from the Riemann surface.

\begin{figure}
\includegraphics[width=.5\textwidth]{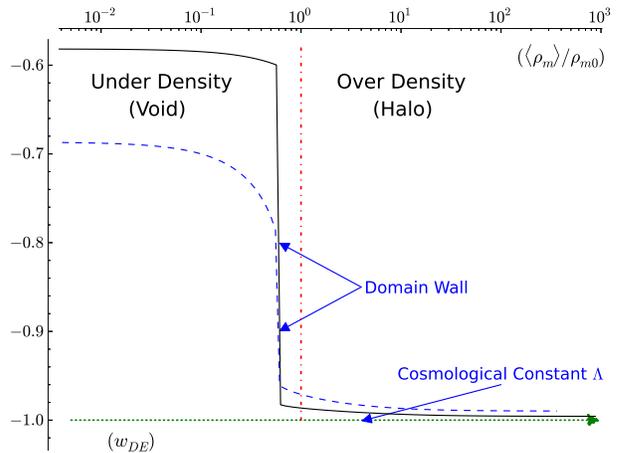}
\caption{ \label{w_de}Plot of EOS of DE  $w_{\rm DE}$  where $\langle\rho_{\rm m}\rangle/\rho_{\rm m0}$ is the mean density of matter interior to a radius $r$
over the mean density today $\rho_{\rm m0}$. {
Blue dashed line (case one) and black solid line (case two).}}
\vspace{-20pt}
\end{figure}

{\bf Dynamical Dark Energy}  Density goes like
$\langle\rho(r)\rangle=\langle{\rho_{\rm m}(r)}\rangle+\rho_{\rm cc}+\rho_{\rm mg}$
 where $\rho_{\rm m}(r)$ is local matter density.
$\rho_{\rm cc}$ is the energy density from the CC.
 We identify $\rho_{\rm mg}(r)$ as the energy density from  massive gravity via the last terms of Eq.~\ref{metric1}.
Upon neglecting radiation, $\langle p\rangle(r)=-\rho_{\rm cc}+p_{\rm mg}(r)$ where $p_{\rm mg}$ is an effective pressure
from massive gravity due to the last terms in Eq.~\ref{metric2}.

Massive gravity naturally leads to inhomogeneous DE.
In Fig.~1, the $\pi$ field responds to changes in the  local density.  As $\pi$ evolves,
so does $\rho_{\rm mg}$ and $p_{\rm mg}$.
The EOS of DE is
$
w_{\rm DE}=\frac{\rho_{\rm cc}+\rho_{\rm mg}}{\langle p\rangle+p_{\rm mg}}.
$
See Fig. 2 for two different cases.  Outside of halos and voids, {%
\bf $w = -0.98\, (-0.95)$}.
As we move into a halo,  $w$ asymptotes towards $-1$.  At the center of a void, {
 $w_{\rm DE}$ jumps to$-0.69\, (-0.58)$. Hence the value of $w_{\rm DE}$
inside a void depends upon the mass of the graviton $\alpha,\beta,\gamma$,\, etc.}

We can also read Fig.~2 as a measure of how the EOS of DE changes with redshift.  At high redshift, the average density
of the universe becomes large compared to the mean density today. The scalar field becomes unimportant which is merely a manifestation of the Vainshtein mechanism. DE is then unimportant for the evolution of the universe.

Recent CMB results from the {\it Planck} satellite \cite{PlanckCosmo} indicate (with $2.5\sigma$ significance)
a nearly $10\%$ lower value of the Hubble constant (H$_0$) compared to values inferred  from the local-Universe and supernova, e.g.~\cite{RiessH}.  
 SNIa systematics and concordance-model uncertainties could account for this difference.
 Our model could also explain the effect with  a significant fraction of low-density void in the local Universe or along the line of sight. 
 With a difference  of $\Delta w_{\rm DE}$ between $w_{\rm DE}^{\txt{voids}}~\approx~-0.7$ in voids and $w_{\rm DE}^{\txt{global}}~\approx~-1$ elsewhere, 
 a higher value of H$_0$ would be inferred than from the 
cosmic microwave background since $\Delta \H/\H_0 \approx 3z\Delta w_{\rm DE}/2w_{\rm DE}^{\txt{global}}$ between the local and global values. The required local under-density of $< \rho_{\rm m0}/2$ 
 appears consistent with recent observational studies, e.g.~\cite{BlomqvistSNvoid, KeenanLocalVoid}. 
This could potentially relieve the tension between the different Hubble constant estimates up to the observed $10\%$ level.

{\bf CONCLUSION}
Massive gravity can naturally give rise to an inhomogenous EOS of DE, which will be 
dependent upon local matter density and scale.  The scale-dependent EOS of dark energy
may reduce the tension between relatively local measurements of dark energy from SNIa and global measurements from the CMB.
Detailed analysis of  voids may be the most direct way to infer the existence of a scale-dependent and density-dependent
EOS of DE since perturbations in matter will lead to perturbations in DE.  We can infer the existence of inhomogeneous DE via dynamics and lensing. 
The motion of galaxies from the geodesic equation constrains $\Phi$ (in the weak-field limit). However, for massless particles, the geodesic equation (lensing) depends upon the lensing potential $\Phi_{\rm L}=1/2(\Phi+\Psi)$.
We can infer the amount of material in and around a void by simply counting galaxies.
It is then possible to infer the EOS of DE in a void and test the predicted large gravitational slip between the metric potentials, which should be achievable in future surveys
\cite{2012arXiv1210.2446K,2012arXiv1211.5966H}. 

{
The {\it Euclid} satellite~\cite{euclid} will measure the stacked void lensing (shear and magnification) signal with S/N $\sim 15$, to derive radial lensing-potential profiles~\cite{2012arXiv1210.2446K}. 
This, combined with the galaxy-density map, should allow for a strong detection/exclusion of the $\sim 50 \%$ weaker lensing potential relative to GR, due to massive gravity. See further also Refs. \cite{2012arXiv1210.2446K,2012arXiv1211.5966H} on e.g. the Baryon Oscillation Spectroscopic Survey (BOSS) and the Large Synoptic Survey Telescope (LSST). 

The EOS of DE will change the growth rate of voids. 
By measuring the shape of voids in redshift space, the Alcock--Paczynski test can reveal local deviations in the expansion rate due to a different equation of state~\cite{2012ApJ...754..109L}. 
For massive gravity, the average equation of state in the universe until today $w_{\rm DE}^{\rm global}\approx-1$. In voids, it is $w_{\rm DE}^{\rm voids}~\sim~-~0.7$, as empirically roughly 50-75$\%$ of void volumes have 
$\rho~<~\rho_{\rm m0}/2$~\cite{2012ApJ...761...44S}. The {\it Euclid} mission could detect a deviation $|\Delta w_{\rm DE}|~>~0.1$ at the $95\%$ level or better~\cite{2012ApJ...754..109L}.  
A deviation $\Delta w_{\rm DE}~=~0.3$ should be highly detectable. For BOSS \cite{boss}, the equivalent limit is $|\Delta w_{\rm DE}|~>~0.25$, allowing for marginal detection. Further, the large-scale-structure power spectrum will be imprinted with a particular length scale derived from the EOS transition, as well as an overall effect on the growth rate. 
Importantly, galaxy surveys also probe the power spectrum and redshift-space distortions, and could potentially  detect local deviations of DE.}  Finally, lensing and polarization of the cosmic microwave background by voids/halos could also be good probes for constraining the model, as well as the integrated Sachs--Wolfe (ISW) effect due to voids/halos~\cite{Inoue:2006fn}.

The domain wall merits further study by potentially affecting lensing and dynamics of  galaxies. With the presence a magnetic field, charged particles could be accelerated
by interacting with the $\pi$ field. As with axions, distinct photon conversion/polarization could take place. Finally, the tension in $\pi$ goes like $\tau\sim\Lambda^3$ by dimensional analysis. 
Matching the metric potentials across the discontinuity in $\pi^\prime$ will give a precise solution for 
the tension in the domain wall.
We should also be able to calculate the wall tension via non-linear field dynamics  and via an analysis of the vacuum structure of $\pi$.
GR implies that they should all give the same solution, which is a curious mathematical connection.

{\bf ACKNOWLEDGEMENTS}
We thank  J.~Khoury for extensive discussion and help.
We also thank P.~Ferreira, W.~Hu, M.~Kamionkowski, E.~M\"ortsell, R.~Rosen, S.~Sj\"ors, and M.~Wyman for discussions. JS and DS were supported by a grant from the ERC, and MS by the Templeton Foundation.

\end{document}